# A Null-model Exhibiting Synchronized Dynamics in Uncoupled Oscillators

Tuhin Chakrabortty[1], Akash Suman[1], Anjali Gupta[2], Varsha Singh[3,#], Manoj Varma[1,4,*]

[1] Center for Nano Science and Engineering, Indian Institute of Science, Bangalore
[2] Center for BioSystems Science And Engineering, Indian Institute of Science, Bangalore
[3] Molecular Reproduction, Development and Genetics, Indian Institute of Science, Bangalore
[4] Robert Bosch Centre for Cyber-physical Systems, Indian Institute of Science, Bangalore

[#] varsha@iisc.ac.in, [*] mvarma@iisc.ac.in

The phenomenon of phase synchronization of oscillatory systems arising out of feedback coupling is ubiquitous across physics and biology. In noisy, complex systems, one generally observes transient epochs of synchronization followed by non-synchronous dynamics. How does one guarantee that the observed transient epochs of synchronization are arising from an underlying feedback mechanism and not from some peculiar statistical properties of the system? This question is particularly important for complex biological systems where the search for a non-existent feedback mechanism may turn out be an enormous waste of resources. In this article, we propose a null model for synchronization motivated by expectations on the dynamical behaviour of biological systems to provide a quantitative measure of the confidence with which one can infer the existence of a feedback mechanism based on observation of transient synchronized behaviour. We demonstrate the application of our null model to the phenomenon of gait synchronization in free-swimming nematodes, C. elegans.

## I. INTRODUCTION

The phenomenon of phase synchronization can be observed in a wide range of physical and biological systems [1] such as mechanically coupled pendulums [2] , firefly strobes, cricket chirping [3], audience applause[4], musical and dance rhythms[5], neuronal activity[6] and respiratory rythms[7]. Here we focus only on the mutual synchronization of autonomous oscillators such as two dancers synchronizing their movements, as opposed to synchronization in oscillators driven by master-clocks, such as the coherent beating of cardiac cells or the circadian rhythms[8, 9]. The phenomenon of synchronization has been extensively analysed using the paradigmatic model of coupled limit cycle oscillators, most notably by Kuramoto [10] . Phase synchronization emerges in this model due to the mutual coupling between oscillators through an instantaneous feedback term with a prescribed strength. The basic Kuramoto model has been extended to include stochasticity [11] , delayed coupling [12], inertia[13] and with generalized coupling functions [14]. In simple physical systems such as two mechanical pendulums connected by a string, it is straightforward to identify the coupling mechanism explicitly. In contrast, biological systems are generally significantly more complex and often, not only the exact coupling mechanism, but also even the number and identity of the oscillators and the toppology of their connected network leading to the observed synchronized phenomenon is not known precisely. For instance, one may not be able to decide between a master clock driven or a mutual entrainment driven phenomenon simply from the observation of synchronized behaviour. Knowledge of exact mechanisms of coupling can have significant benefits such as allowing the prevention of synchronization of spontaneous neural activity during epileptic seizures[15, 16] .

As mentioned before, there is a vast literature describing models which exhibit phase synchronization incorporating effects of stochasticity, finite signal propagation time, inertia and so on. However, a robust understanding of synchronization phenomenon requires a null model which quantifies the probability that the observed synchronization phenomenon arose out of pure statistical coincidence without any coupling mechanism. In a noisy complex system, such as the one described in this article, one generally only observes transient epochs of synchronization followed by noise induced de-synchrony [17]. In such situations, how does one ensure that the observed transient synchronization events are indeed the result of an underlying feedback (or a master-clock) mechanism and not merely a statistical coincidence? One requires an appropriate null model to guard against inference of non-existent feedback mechanisms based on observation of such transient synchronization events. Having such a null model would save the effort involved in searching for a non-existent feedback mechanism in a complex system. This paper considers the problem of constructing an appropriate null model which will serve as the baseline against which one can assess the extent to which experiments suggest the existence of a feedback mechanism. Despite the large literature available on models describing synchronization, we are not aware of any study which characterizes a null model in detail.

We apply our null model to a simple biological synchronization phenomenon described by Yuan et al. involving gait synchronization in pairs of swimming nematodes (*C. elegans*) confined in microfluidic channels [18]. In contrast to Yuan et al. we perform gait synchronization experiments on free-swimming nematodes and show that a null model without feedback also produces statistically identical patterns of transient synchronization observed in the experimental system. The key idea of our null model is that most biological systems (as well many non-biological real-world systems) are likely to have a finite response time, which we refer to as the persistence time. Then, we would expect an accidental locking of phases at any instant of tme to persist for a time comparable to the persistence time-scale. Thus, an experimental observer would note several transient synchronization events despite the existence of any feedback or coupling between the oscillators. These observations will in general be similar to those occurring in noisy systems with real feedback. Our null model establishes bounds on the probability of occurrence of coincidental synchronization. In particular, we show that unambiguous determination of the existence of feedback requires the experimenter to be aware of the system response time (persistence time-scale). Moreover, the probability of coincidental synchronization decays exponentially with the number of oscillators. Therefore observation of synchronized behaviour in larger systems is a strong indication of the presence of a feedback mechanism. For instance, in our experimental system, observation of large number of worms (up to 7) swimming synchronously establishes the role of a feedback mechanism as suggested by Yuan et al., because the probability of occurrence of such an event is negligible in the null model. Thus, we establish quantitative bounds for inferring the existence of feedback from observation of transient synchronization events. Finally, we show that incorporating a feedback term into the null model recovers some of the results described in past literature, such as simultaneous phase and frequency locking in the presence of inertia [19].

In the subsequent sections we show the results from our experiments, describe the null model in detail and illustrate the utility of the null model in reliably establishing the role of a feedback mechanism in gait synchronization in free-swimming nematodes.

## II. RESULTS

### A. Gait synchronization in free-swimming nematodes

We observed free-swimming nematodes (also referred to as worms in this article) in a droplet placed under a stereo microscope [*details in the methods section at the end of this article and SI text section (1)*]. The free-swimming nematodes were sometimes observed to synchronize their gaits while in close proximity of each other. *[Fig 1(a) and SI video1]*. We analysed 117 pairs of worms with synchronised gaits swimming less than 0.2 mm apart from each other for at least 5 seconds and determined the distribution of synchronization times ($n_s$) [Fig 1 (b)]. Synchronization events were identified by calculating the relative phase difference between their respective swim strokes as descried in the methods section *[See SI video 2]*. In Yuan et al., probability of gait synchronization between two worms increases as they come closer. This observation led them to propose a steric hindrance led feedback mechanism resulting in the observed gait synchronization. As they performed their experiments in a confined microfluidic channel, majority of the worm pairs in close proximity synchronized their gaits. Contrary to this, in our case, pairs of worms didn't synchronize despite their close proximity to each other in more than 50% of the cases. Taken in isolation, this points out to the strong influence of experimental geometry in the observed behavior which should caution against the application of such findings in more general settings. Another point worth mentioning is that gait synchronization can be observed even when worm pairs are so far apart to rule out any realistic physical coupling mechanism as seen in Fig. 1 (d) and SI video 3. Comparison of Fig. 1 (c) and (d) shows that relative phase difference is not a robust measure of synchronization as zero relative phase difference can occur due to statistical coincidence as in the case of Fig. 1 (d). This fact is evident in the experiments of Yuan et al. as well, indicated by the occurance of synchronization parameter close to unity even for worms very far apart [Fig 2b in Yaun et al.]. Examples of such spurious synchronization events clearly motivates the need for a null model to account for such instances.

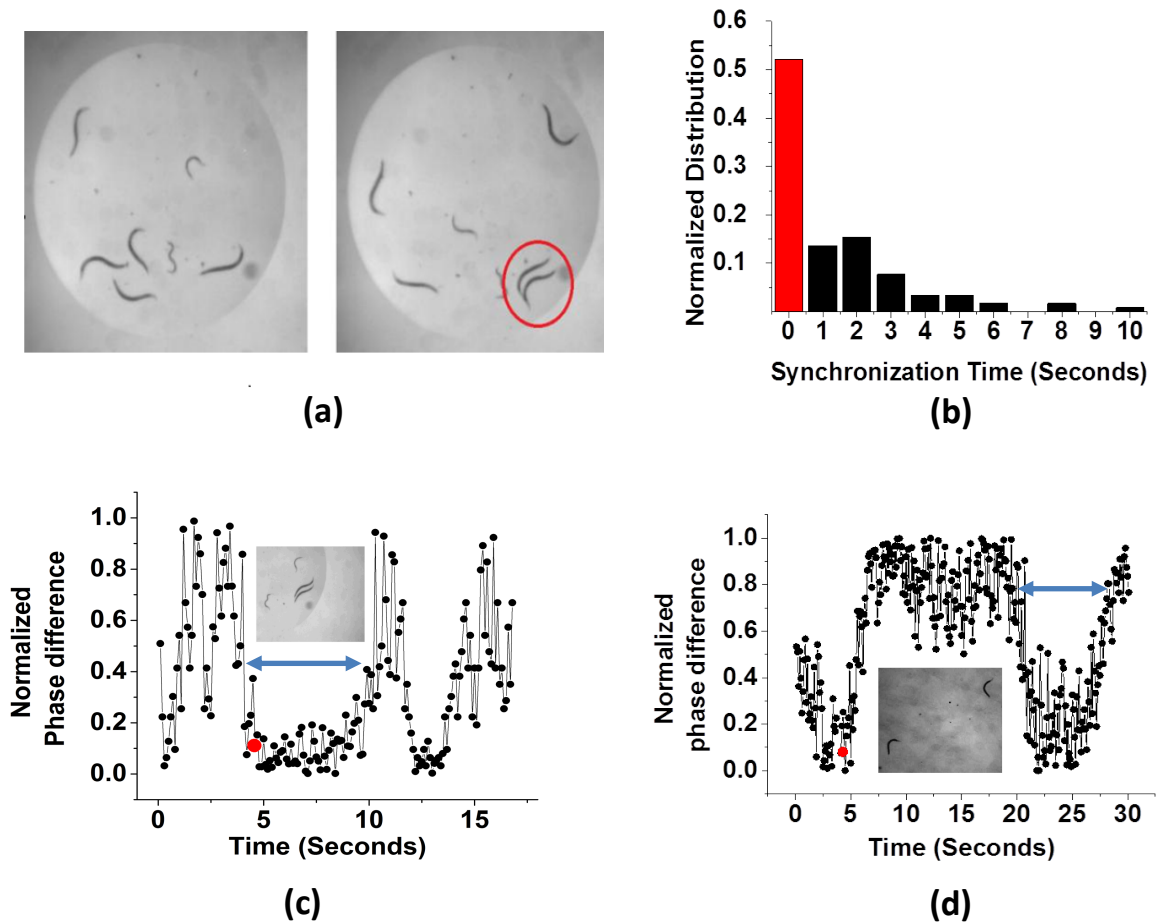

**Figure 1: (Color online)** *Experimental results (a) Worms swimming freely in a droplet and a pair of worms swimming synchronously, (b) Normalized distribution of synchronization time for 117 pairs of worms swimming close to each other. The 'red bar' indicates the fraction of worm pairs, which were not synchronized. (c) and (d) show the variation of phase difference for pair of worms swimming close to each other and far from each other respectively. The 'red marker' shows the phase difference at the time frame shown in the image. The 'valley' (shown by the The 'double sided arrow') in the phase difference plot represents a synchronization event. It can be observed that, for free-swimming worms, the duration of the synchronization event can be similar irrespective of the distance between the worms. .*

## B. Description of the null model

We construct our null model based on two factors which we believe to be important in biological systems (as well as many non-biological real-world systems), namely the effect of stochasticity as well as a finite persistence or correlation time-scale. Such a persistence time scale can arise due to the inertia of the oscillator leading to finite response time. Thus, parameters such as the beat-period of the worm cannot instantaneously change. In accordance with our experimental method (See methods section for details), we use a discrete model to describe the beat-period of the worms. For $N$ worms, labelled by index $i$, the beat-period $\beta$ in the $n^{th}$ cycle is written as

$$\beta_n^{(i)} = \bar{\beta}(1 - \lambda\alpha) + \alpha\lambda\beta_{n-1}^{(i)} + \lambda\xi_n^{(i)} \quad (1)$$

This equation is a first order autoregressive (AR) model [20] for the beat-period where $\bar{\beta}$ is the mean beat-period and $\xi$ is a delta-correlated Gaussian white noise term with $\langle \xi_n^{(i)} \rangle = 0$ and $\langle \xi_m^{(i)} \xi_n^{(j)} \rangle = 1$. The product of parameters $\lambda$ and $\alpha$ controls the correlation or persistence time-scale, going from a $\delta$-correlated system (i.e. zero persistence time) at $\lambda\alpha = 0$ to a system with significant persistent time as $\lambda\alpha$ approaches 1. The magnitude of the product $\alpha\lambda$ needs to be less than one to ensure stationarity of the time series [20]. As we will show later, the form of Eq (1) is motivated by the Langevin equation with inertia.

The stationary variance of an AR(1) process is

$$\sigma_{\beta,d}^2 = \frac{\lambda^2}{1 - \alpha^2\lambda^2}$$

Here the subscript d denotes the discrete process represented by Eq (1).

and the persistence time-scale ($\nu_c$) for the AR model is given by *[See SI section 3(a) for details]*

$$\nu_c = -\frac{1}{\log(\alpha\lambda)}$$

We take this AR model as a discrete representation of a continuous Langevin model for $\beta(t)$ given by

$$m\dot{\beta}^{(i)}(t) = -\gamma\beta^{(i)}(t) + \xi^{(i)}(t) \quad (2)$$

where:
$$\xi^{(i)} = \mathcal{N}(0,1)$$

The stationary variance of the Langevin process is

$$\sigma_{\beta,c}^2 = \frac{1}{2\gamma m}$$

and the persistence time-scale is given by

$$\tau_c = m/\gamma$$

In order for the AR model in Eq. (1) to be a physically equivalent description of the Langevin process described by Eq. (2), we demand the stationary mean, variance and the correlation time scale of the two models to be equal. Equivalence of the AR model and the Langevin model leads to,

$$\alpha\lambda = e^{-\frac{\gamma}{m}} \quad (3)$$

Equation (3) allows us to interpret the product term $\alpha\lambda$ of the discrete model in terms of inertia of an equivalent continuous model. A system with larger inertia will have a larger value of $\alpha\lambda$ and consequently longer persistence time-scale. Further, imposing physical equivalence leads to a unique solution of $\lambda$ and $\alpha$ of the discrete AR(1) model in terms of the mass and damping factor of the Langevin equation. *[See SI section 3(a)]*. Increasing mass leads to scaling down of the stochastic noise term in the Langevin equation (Eq (2)). The parameter $\lambda$ is inversely related to the mass and leads to a suppression of the noise term in the discrete model in an analogous way.

We define an equivalent "phase" *[SI section 3(a) for details]* of the oscillator described by Eq. (1) as,

$$\phi_n = \phi_{n-1} + \beta_n$$

Two oscillators are defined to be phase synchronized when the phase difference between them is less than a pre-defined threshold $\delta$. One could interpret $\delta$ as the precision of the experimental measurement technique. For example, in a video analysis of swimming gaits of nematods, $\delta$ would be set by the frame rate, magnification and pixel intensity noise level in the video.

Following the derivation presented in SI text section 3(a), for timescales shorter than the persistence time, the probability of seeing a synchronization event of $n_s$ cycles for two oscillators, during an observation period of $n_{obs}$ can be obtained as

$$p(n_s, 2 \mid \alpha\lambda \to 1) = \left[1 - \text{erf}\left(\frac{(\bar{\beta}-\delta)}{2\sigma_\beta}\sqrt{\frac{|\log(\alpha\lambda)|}{n_{obs}}}\right)\right]\text{erf}\left(\frac{\delta}{\sigma_\beta n_s \sqrt{2}}\right) \quad (3)$$

Where $\bar{\beta}$ and $\sigma_\beta^2$ are the mean and variance of the beat-period of a single oscillator.

Similarly for a $\delta$-correlated system, the probability of seeing a synchronization event is given by

$$p(n_s, 2 \mid \alpha\lambda \to 0) = \left[1 - \text{erf}\left(\frac{(\bar{\beta}-\delta)}{\sigma_\beta\sqrt{2n_{obs}}}\right)\right]\left[\text{erf}\left(\frac{\delta}{\sigma_\beta\sqrt{2}}\right)\right]^{n_s} \quad (4)$$

For N oscillators to be coincidentally synchronized (i.e. without feedback), we require each pairwise phase difference to be within $\delta$ and this leads to an exponentially decaying probability of observation of coincidental synchronization as,

$$p(n_s, N) = [p(n_s, 2)]^{\frac{N(N-1)}{2}} \quad (5)$$

To incorporate feedback, we modify the model in Eq. (1) to

$$\beta_n^{(i)} = \bar{\beta}(1 - \lambda\alpha) + \lambda\alpha\beta_{n-1}^{(i)} + K\left(\bar{\phi}_n - \phi_n^{(i)}\right) + \lambda\xi_n^{(i)} \quad (6)$$

where $\bar{\phi}_n = \frac{1}{N}\sum \phi_n^{(i)}$ and K is the feedback coupling constant.
while the expression for phase $\left(\phi_n^{(i)}\right)$ remaining the same.

$$\phi_n^{(i)} = \phi_{n-1}^{(i)} + \beta_n^{(i)}$$

For the two oscillator case, the equation (7) becomes

$$\beta_n = \alpha\lambda\beta_{n-1} + \lambda\xi_n - K\lambda\phi_n \quad (7)$$
$$\phi_n = \phi_{n-1} + \beta_n$$

Where $\phi_n$ and $\beta_n$ are $\left|\phi_n^{(1)} - \phi_n^{(2)}\right|$ and $\left|\beta_n^{(1)} - \beta_n^{(2)}\right|$ respectively.

Equation (8) admits a stationary solution for $\beta$ and $\phi$ for sufficiently large feedback. *[Please refer to SI section 3(b) for more details]*. The stationary variance of $\beta$ and $\phi$, namely, $\sigma_\beta^2$ and $\sigma_\phi^2$ are then obtained as,

$$\sigma_\beta^2 = \frac{2\lambda^2}{(K\lambda+1)(K\lambda+2) - 2\alpha^2\lambda^2}$$

and

$$\sigma_\phi^2 = \frac{\sigma_\beta^2(1 + K\lambda)}{2K\lambda}$$

In this case, the probability of seeing a synchronous state for two oscillators can be obtained as *[SI section 3(b)]*

$$p(n_s, 2) = \left[\text{erf}\left(\frac{\delta}{\sqrt{2}\sigma_\phi}\right)\right]^{n_s} \quad (8)$$

We checked the validity of this model using numerical simulations as well as the experimental data of gait synchronization in free-swimmiing nematodes described earlier.

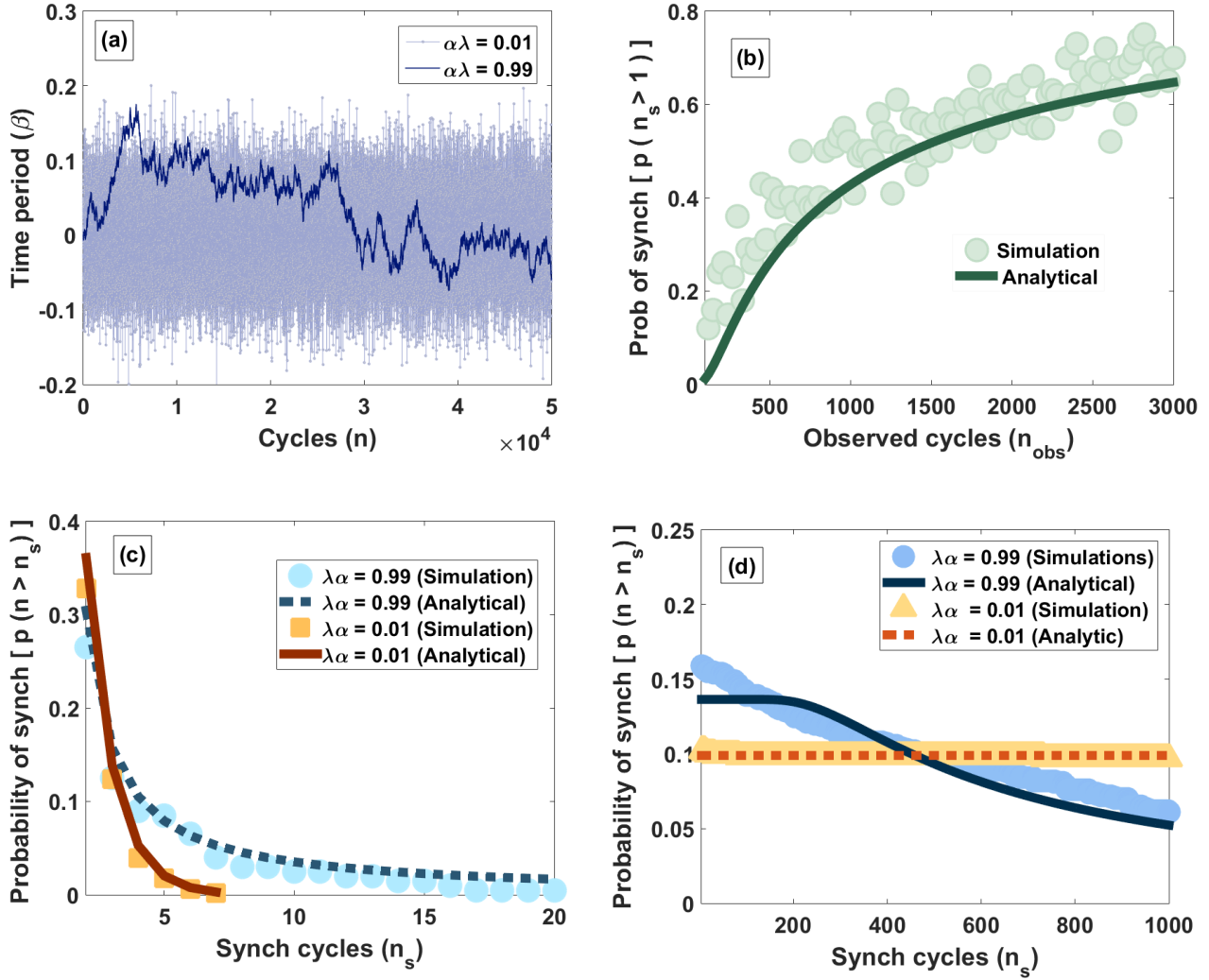

**Figure 2: (Color online)** *Characteristics of coincidental synchronization:* ***(a)*** *Effect of $\lambda\alpha$ on the time period.* ***(b)*** *Effect of $n_{obs}$ [Parameters: $\bar{\beta} = 0.5$, $\delta_\beta = 0.05$, $\alpha\lambda = 0.99$, $\sigma_\beta = 0.001$* ***(c)&(d)*** *Effect of $\alpha\lambda$ at different noise levels.* *[**(c)** Parameters: $\sigma_\beta = 0.1$, $\bar{\beta} = 0.5$, $\delta_\beta = 0.05$, $n_{obs} = 1000$* ***(d)*** *Parameters: $\sigma_\beta = 0.0001$, $\bar{\beta} = 0.5$, $\delta_\beta = 0.05$, $n_{obs} = 1000$. ]*

### C. Comparison of numerical simulations with theoretical model

Figure 2 presents the key aspects of the AR model described by Eq. (1). Fig. 2 (a) shows the effect of increasing $\lambda\alpha$, namely an increase in the correlation (persistence) time-scale as mentioned in the previous section *[Refer SI text section 3(a) for more details]*, [20]. The null model represented by Eq. (1) was able to capture the scaling of the probability of coincidental synchronization (Fig. 2b) and the distribution of synchronization times (Fig 2c and 2d) correctly, indicated by the close agreement with numerical simulations. As seen in Fig. 2c, for a relatively high level of noise (complete set of parameters mentioned in caption), as expected from our discussion in the previous section, we see that longer persistence times (indicated by $\alpha\lambda$ approaching unity) leads to a greater probablity of observing long synchronization epochs comparable to the persistence time-scale.

However, the AR model described by Eq. (1) possesses a non-trivial behavior with noise. An indication of this is shown in Fig. 2d, where longer persistence time-scales lead to reduced probability of observing long synchronization epochs. In fact, the simulations as well as the theoretical model indicate that it is possible to observe synchronization during the entire observation period for small $\alpha\lambda$. This is because, for low noise levels, the fraction of events with initial phase difference lying within the threshold will tend to remain so for systems close to the delta-correlated limit. However, when the system deviates from a delta-correlated behavior, long-term drifts lead to desynchronization from the initially synchronized state. Even in this case, it can be seen that the probability of short-term synchronization epochs is greater for systems with longer persistence time-scale.

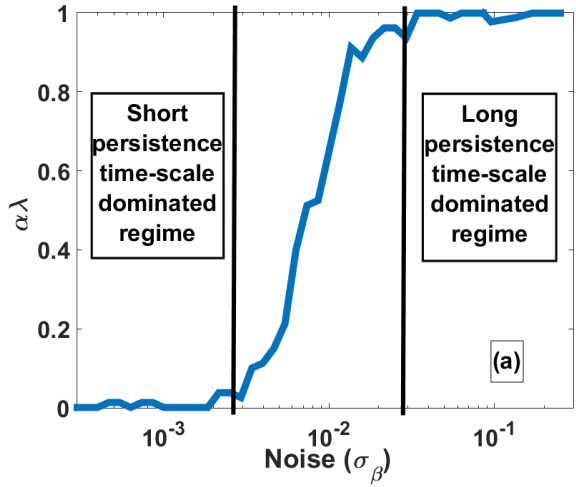
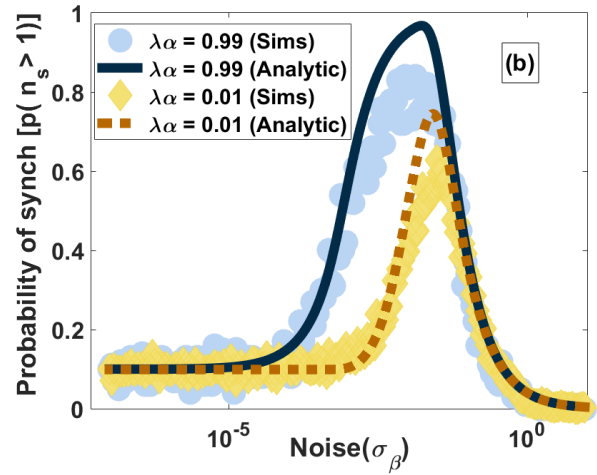

**Figure 3. (Color online)** *(a) Effect of $\alpha\lambda$ on observation of longer synchronization epoch as a function of noise $(\sigma_\beta)$. (b) Probability of observion of synchronization epoch for different values of $\alpha\lambda$ as a function of noise $(\sigma_\beta)$. [Parameters: $\bar{\beta}$ = 0.5, $\delta_\beta$ = 0.05, $n_{obs}$ = 1000.]*

The complete synchronization behavior of this system with noise is shown in Fig. 3. First, we ask the question whether a shorter or a longer persistence time leads to longer synchronization epochs as a function of noise. In Fig. 3a, we see that there is a transition from a regime favoring short persistence time-scales to one favoring long persistence time-scales as noise is increased. Another interesting aspect of our coincidental synchronization model, is the noise enhanced likelihood of this effect, analogous to systems displaying stochastic resonance. In other words coincidental synchronization is more likely at an optimal noise level as shown in Fig. 3b. The origin of an optimal noise level can be understood in the following manner. In order to achieve a synchronized epoch with length $t_s$ starting from an arbitrary phase difference, firstly the relative phase difference should diffuse to within the threshold $\delta$ (phase-locked state) and then should continue to remain within the threshold for at least $t_s$. Reaching the phase locked state is facilitated by larger noise levels whereas staying within that state is not. These two counteracting effects lead to the formation of an optimal noise level which maximizes the probability of coincidental synchronization as shown in Fig. 3b. *[See SI section 3a for details]*

Incorporating feedback into the null model results in significantly longer epochs of synchronization, relative to the null model, with increasing $\phi_n$. For a relatively small feedback, it becomes extremely difficult to predict the system coupling constant as expected (Fig. 4a). Figure 4b shows the effect of feedback on the stationarity of behaviour analytically because the system does not attain stationarity within the observation window ($n_{obs}$) considered. *[Please see SI text section 3(b) for details]*. Incorporating feedback allows us to recover a previosuly described result [19], namely simultaneous synchronization of phase and frequency in systems with inertia. A notable consequence of incroporating feedback is that the probability of multiple oscillators synchronizing becomes significant, unlike the null model where the probability of multiple oscillators synchronizing drops rapidly. For larger number of oscillators, it is more appropriate to describe synchronization in terms of the order parameter [19, 21]. We define the order parameters for phase $r_\phi$ and beat-period $r_\beta$

$$r_\phi(n) = \frac{|\sum_k e^{i\Delta\phi_k(n)}|}{N-1}$$

and

$$r_\beta(n) = \frac{|\sum_k e^{i\Delta\beta_k(n)}|}{N-1}$$

Where N is the number of oscillators and $\Delta\phi_k(n)$ and $\Delta\beta_k(n)$ are the phase difference and time period difference respecively between $k^{th}$ pair of oscillators at $n^{th}$ cycle. Figures 4c and 4d show that both order parameters approach unity as time progresses indicating simultaneous frequency and phase synchnronization. In contrast, neither of the order parameters converge to unity for the null model indicating that the probability of multi-oscillator synchronization without feedback is very low

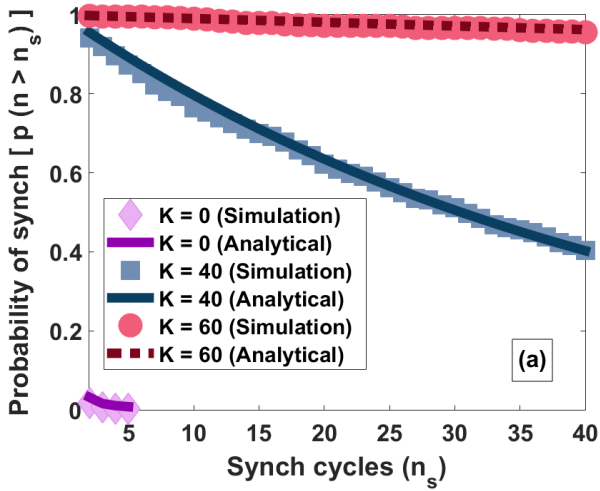
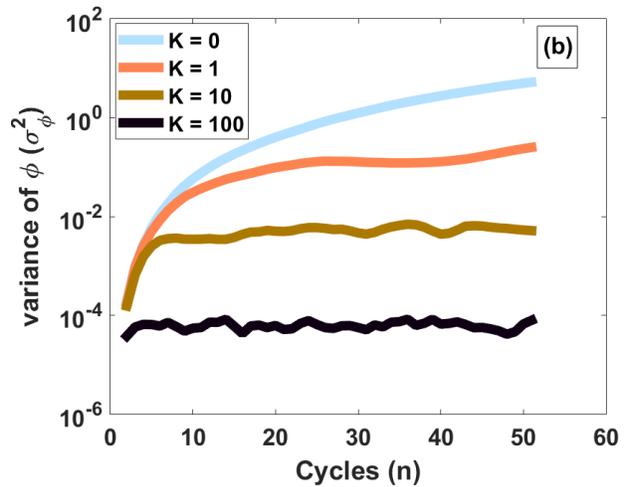
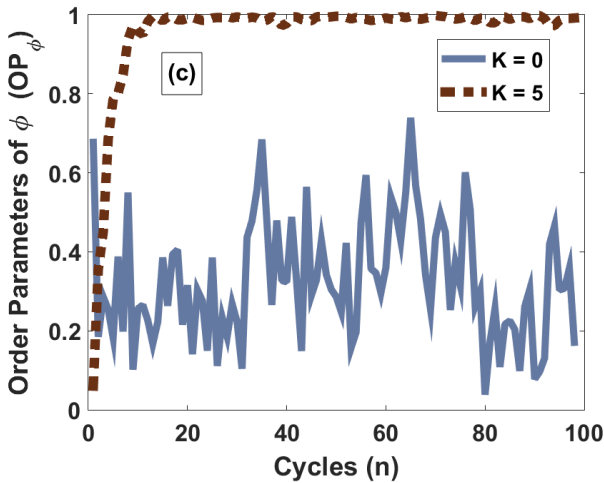
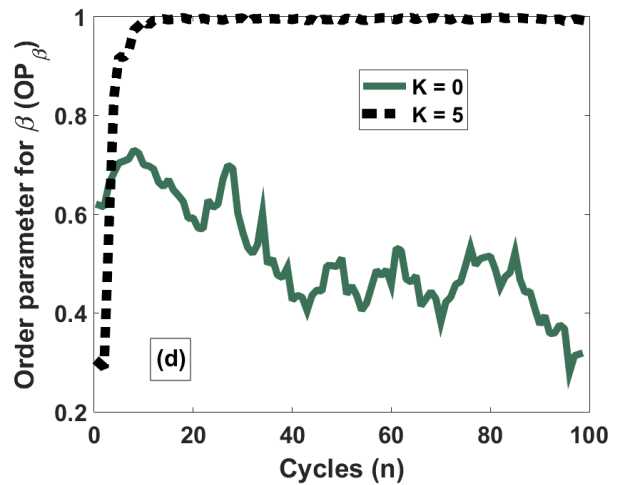

**Figure 4**: **(Color online)** *Effect of feedback*: (a) Comparison of $p(n_s,2)$ for feedback and null model for large K. Parameters: $\alpha\lambda = 0.99$; $\sigma_\beta = 1$; $\delta_\beta = 0.05$, $\bar{\beta} = 0.5$, $n_{obs} = 100$ (b) Effect of K on stationarity of $\sigma_\phi^2$. [parameters: $\alpha\lambda = 0.99$; $\sigma_\beta = 0.1$; $\delta_\beta = 0.05$, $\bar{\beta} = 0.5$ (c)&(d) Evolution of order parameters for simulated phases ($\phi$) and Time period ($\beta$) respectively. [Parameters: $\alpha\lambda = 0.99$, $\bar{\beta} = 0.5$, $\sigma_\beta = 1$, $N_{osc} = 10$

Having convinced ourselves of the validity and utility of the null model and the feedback model, we are finally in a position to compare the experimental data against these two models. In order to compare the experimental observation with predictions of the null model, we first extracted the AR model parameters from the experimental data as described in the methods section. The data of two-worm synchronization is not sufficient to decide between the two models as one can obtain reasonable fits to the data with either one of them (Fig. 5a). However, the multiple-worm synchronization data, where we see upto 7 worms synchronizing their gaits *[See SI video 4]* cannot be explained by the null model (Fig. 5b) and therefore strongly supports the proposal by Yuan et al. of the existence of a feedback mechanism, specifically the steric hindrance one proposed in their work. The steric hinderence based feedback is further supported by the fact that multiple worm synchronization is observed during drying up of water droplet containing the worms.

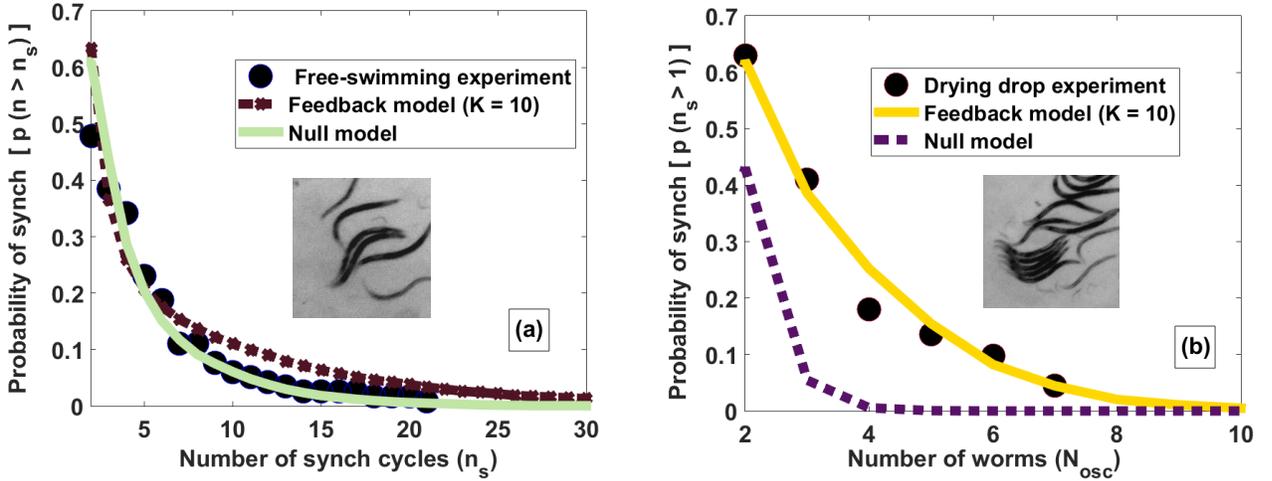

**Figure 5: (Color online)** *Comparison of experimental data with models*: (a) Distribution of synchronization time for free swimming worm pairs is compared with both the models. [Parameters used for the models, determined from the experimental data, are $\alpha\lambda = 0.97$, $\bar{\beta} = 0.52$, $\sigma_\beta = 0.01979$, $n_{obs} = 100$. (b) Probability of observing synchronization of $N_{osc}$ worms in a drying droplet is compared with the feedback and null model. The observation period ($n_{obs}$) of drying droplet is much lesser than the observation period of the free swimming worms. [Parameters used for the models, (determined from the experimental data), are $\alpha\lambda = 0.97$, $\bar{\beta} = 0.52$, $\sigma_\beta = 0.01979$, $n_{obs} = 30$. K (=10) is determined by fitting the experimental data from the drying droplet with the feedback model]

### III. Discussion

In their paper, Yuan et al. [18] reported that, worm pairs confined in a microfluidic channel synchronize their gaits in close proximity. They proposed steric hindrance as the causative feedback mechanism leading to synchronization. The tightly confined space of a microfluidic channel introduces a very different environment from the natural free-swimming environment of the worm which prompted us to perform free-swimming experiments. As we have seen in the case of free swimming worms, it is possible to observe an apparently synchronous state even for worms far apart to rule out any effective physical feedback mechanism. Therefore, a rigorously supported conclusion of synchronized behaviour of the worms must account for the random occurrence of transient synchronization events likely in this system. The beat-period of the worm has a persistence time-scale arising out of finite nerve signal propagation, elastic constants of the muscles etc. which leads us to model the beat-period as in Eq(1). We see that such a system is capable of producing synchronized epochs of duration comparable to the persistence time-scale of the system for two-worm synchronization. However, as we saw, the probability of such random synchronization events decays very sharply as the number of worms (oscillators) increase. Further, the probability of observing a synchronization event significantly longer than the persistence time also decays rapidly. Our analysis therefore suggests that a strong evidence for demonstrating synchronization in systems with finite response time is to explicitly measure the persistence time-scale and check if the observed synchronization duration is significantly longer than this time-scale. The null model discussed here provides a quantitative means to perform this check. An equally effective method, in light of the analysis presented here, is to perform multi-oscillator experiments. The null model predicts very low probability for multi-oscillator synchronization. Therefore, observation of multi-oscillator synchronization in experiments is a strong evidence for the existence of feedback. In view of this, two-worm experiments are not strong enough for concluding the existence of feedback. Synchronization of gaits of worms swimming in thin liquid films (as the liquid is absorbed into the agar medium) provides a strong evidence that when confined, either by the drying liquid film or in a microfluidic channel as in Yuan et al., the worms enter a regime which involves a feedback mechanism, very likely the steric hindrance mechanism proposed by Yuan et al. So while our data and conclusions do not conflict with the previous results, our analysis presents a more qualified picture of the phenomenon. In addition, the mathematical framework presented here provides a good null model to serve as baseline in determining the confidence to which a feedback mechanism may be expected based on experimental data suffering from limited sampling ($n_{obs}$) and measurement precision ($\delta_\beta$). The characteristics of random synchronization is the lack of a stationary phase characterized by a stationary order parameter approaching unity for large enough feedback.

## IV. Methods

### A. Observation of gait synchronization of free-swimming worms

In the experiment, a 40-50 $\mu$l drop of DI water containing 8-10 worms was put on an agar surface using a micro pipette. The worms were allowed to swim freely till the water droplet dried (typically 1-2 mins). The dynamics of the swim strokes of the nematodes in the droplet was recorderd using a ThorLabs® camera mounted on a 5x stereo microscope. This experiment was repeated 75 times with different set of worms. Worm pairs were selected at random and were analyzed using custom codes in MATLAB® to calculate the instantaneous phase difference ($\Delta\phi$). Out of all the pairs analyzed, 117 pairs were selected which were swimming less than 0.2 mm apart from each other for at least 5 seconds. The worm pairs were defined to be synchronized if $\Delta\phi$ was less than 0.2. A histogram of synchronization times was plotted for 117 pairs. [Fig 1(b)] We also conducted the experiment for worms at different starvation conditions but did not observe any significant effect of starvation on the beat-period. *[Refer SI section (1) for details about experimental set up and analysis method].*

### B. Numerical simulations

We used Monte Carlo technique to simulate the frequency and phase of the nematodes. Time series of beat-period ($\beta$) of $N$ worms were generated using both null model (Eq. (1)) and the feedback model (Eq. (7)). Instantaneous phases ($\phi$) of each worm were calculated from the time periods. For the case, $N = 2$, phase difference ($\Delta\phi$) was calculated for each time instant and the intervals were noted for which $\Delta\phi < \delta_\phi$. For the case $N > 2$, order parameters were defined as described in Eq(10).

### C. Extracting model parameters from experimental data

To extract the AR(1) parameters from the swimming nematodes, we conducted single worm experiments. The beat periods of the worms were calculated using in-house image processing codes in MATLAB. (*Refer SI section 1 for details*). The time series of beat periods of a single worm was obtained. We then applied the Yule-Walker method [20] on the time series data to extract the AR(1) parameters. The process was repeated for several worms and the average of the parameters were calculated over data from 23 worms. (*Refer SI secton 2 for details*)